\documentclass[conference,compsoc]{IEEEtran}

\ifCLASSOPTIONcompsoc
  \usepackage[nocompress]{cite}
\else
  \usepackage{cite}
\fi
\usepackage{graphicx}
\usepackage{booktabs}
\usepackage{enumitem}
\usepackage{makecell}
\usepackage{tabularx}
\usepackage{xcolor}
\usepackage{xurl}
\usepackage{tikz}
\usepackage{xstring}
\usepackage{hyperref}
\usetikzlibrary{arrows.meta,fit,positioning}
\hypersetup{hypertexnames=false}

\title{Aftershock: Yesterday’s Linux Kernel Bugs, Tomorrow’s Android Exploits}
\author{
\IEEEauthorblockN{Kareem Shehada}
\IEEEauthorblockA{University of California, Riverside\\
ksheh002@ucr.edu}
\and
\IEEEauthorblockN{Zhiyun Qian}
\IEEEauthorblockA{University of California, Riverside\\
zhiyunq@ucr.edu}}

\newcommand{\code}[1]{\texttt{\detokenize{#1}}}

\newcommand{\PP}[1]{
\vspace{2px}
\noindent{\bf \IfEndWith{#1}{.}{#1}{#1.}}
}

\begin{document}

\maketitle

\begin{abstract}
Android kernels have long suffered from slow patch propagation from Linux upstream, with fixes often reaching devices far later than conventional Linux distributions. Until recently, this lag was largely a patch-management problem. With rapidly improving LLM agents, it is becoming an exploitation problem: public upstream patches can now serve as blueprints for building working n-day exploits before downstream devices are fixed.

We test this threat end to end by building an automated pipeline that screens upstream Linux patches, reproduces memory-safety vulnerabilities on Android targets, and drives an LLM agent to construct local privilege escalation exploits from an untrusted application. Across 152 candidate CVEs reported over 33 months, we reproduce 38 vulnerabilities and generate 17 independently verified exploits under production defenses. Exploitation takes a median of 14.3 agent-hours and \$321 in API cost, with 90\% median reliability; 12 of 14 trials succeed on Pixel 9 and Pixel 10 devices. Notably, the pipeline relies only on publicly available, off-the-shelf models.

Surprisingly, only 3 of the 17 exploited CVEs appear in Android Security Bulletins, suggesting limited visibility into which upstream fixes address vulnerabilities that affect Android kernels. For these 17 CVEs, the corresponding upstream patches reach Pixel devices with a median delay of 181 days. Together, these results reveal a stark asymmetry: attackers can increasingly automate exploitation in hours, while defenders still identify and deploy fixes on the timescale of months.

\end{abstract}

\IEEEpeerreviewmaketitle

\section{Introduction}
\label{sec:introduction}

An upstream Linux security fix can be public for weeks or months before it
appears in Android firmware. Fixes typically pass through Linux stable releases,
Android Common Kernel (ACK) branches, and vendor firmware builds before users
can install them~\cite{zhang2021androidpatch,li2024investigation}. In the
meantime, the patch identifies the affected code and helps an attacker
determine which downstream releases still contain the
defect~\cite{maar2025doom,li2026they}. As large language model (LLM) agents
take on work that once required a kernel exploitation specialist, the
practical significance of this \emph{n-day exposure window} depends on what
those agents can accomplish while a fix is still making its way downstream.

Recent benchmarks show that agents can exploit real software vulnerabilities.
ExploitGym~\cite{wang2026exploitgymaiagentsturn} and ExploitBench~\cite{lee2026exploitbenchcapabilityladderbenchmark} evaluate frontier models' offensive capabilities using vulnerabilities in the V8 JavaScript engine and the Linux kernel. 
On Android, a public Linux fix leaves open whether an ordinary app can reach
the defect on a downstream kernel. Establishing local privilege escalation
(LPE) further requires demonstrating that an app starting in the
\code{untrusted_app} SELinux domain can gain elevated privileges with the
target's system-call restrictions and kernel mitigations in place.

We build and evaluate an end-to-end pipeline for generating exploits for Android kernels using upstream Linux kernel n-day vulnerabilities. The pipeline takes public upstream Linux
patches as input and uses publicly available frontier models. It screens for
memory-safety bugs relevant to Android, identifies affected Android Common Kernel targets,
and tests whether an ordinary app can reproduce the bug on a Cuttlefish
virtual device. No pre-built reproducer is supplied. For reproduced cases,
the pipeline evaluates autonomous exploit generation and independently tests
each submitted APK on a freshly booted instance with production mitigations
enabled and no interactive agent access.
Furthermore, the pipeline lets us examine both the outcome of each attempt and the process
that produced it. We measure how often agents achieve LPE, how much time and money it costs, and how reliably the exploits work on repeat trials.
We also test whether they can be adapted to other targets, including real Android phones. 

Of the 152 candidate CVEs that enter runtime reproduction, the pipeline
reproduces 38 from the app sandbox and yields independently verified LPEs for
17. From initial reproduction to verification, successful cases require a
median of 14.3~agent-hours and a median of \$321.07 in model API costs.
Across 170 fresh-instance trials, the median per-exploit success rate is
90\%. We also evaluate adaptation to physical hardware, where 12 of 14
executed trials succeed on Google Pixel~9 and Pixel~10 devices.

For the vulnerabilities with verified exploits, we then trace the upstream
fixes through monthly Generic Kernel Image (GKI) branches and Pixel factory
firmware. Among the eight CVE--device pairs patched during the study, the
first fixed Pixel release appears a median of 181~days after the upstream
fix. On their release dates, 64.1\% of the evaluated Pixel factory images
contain at least one studied vulnerability whose CVE was already public.
Only three of the 17 exploited CVEs appear in the Android Security Bulletin
during the study period. These results expose a disparity between hours of
agent effort in successful cases and months of patch delivery for affected
devices. They also present an opportunity for maintainers. As our pipeline
demonstrates, the automated reproduction and exploit generation can quickly establish the significance of an upstream bug and its applicability to downstream Android devices.

Each verified exploit gives us a glimpse of the kernel exploitation expertise an agent can bring to bear from inside the Android app sandbox. 
Many techniques in the Linux kernel exploitation literature rely on system
calls or kernel interfaces that Android denies to ordinary apps. We examine
the generated exploit code to understand how the agents achieved privilege
escalation from the more restrictive sandbox. 
Across the 17 verified exploits, we find that agents draw on established
Android kernel exploitation strategies and techniques. 
Interestingly, we also observed novel strategies and techniques, something that OpenAI reported without providing any technical details in the context of Linux kernel exploits~\cite{openai_lpe}.

Our study makes four contributions.
\begin{itemize}[leftmargin=*]
  \item \textbf{An end-to-end measurement pipeline for Android kernel n-days.}
  We build a pipeline that connects upstream patch screening and app-context
  reproduction to independent verification of Android kernel LPEs. Of the
  152 candidate CVEs selected for runtime reproduction, we confirm that 38
  are reachable from an ordinary app.

  \item \textbf{Reliability and adaptation across targets.}
  We measure per-exploit reliability across 170 fresh-instance trials and
  evaluate adaptation across kernel revisions and from Cuttlefish to physical
  devices. We verify privilege escalation in 12 of 14 trials (85.7\%)
  executed on Pixel~9 and Pixel~10 devices.

  \item \textbf{Downstream patch propagation and firmware exposure.}
  We trace upstream fixes to GKI branches and Pixel firmware. The median GKI delay was 21.5~days for 201 patched CVE--branch pairs and at least 71.4~days when unpatched pairs are included. Across eight patched CVE--device pairs, fixes took a median of 181~days to reach Pixel~9 and Pixel~10 firmware.

  \item \textbf{Strategies and techniques in agent-generated exploits.}
  We analyze the generated code for all 17 verified exploits and compare
  their strategies and techniques with state-of-the-art Linux kernel exploits.
  We identify how established techniques are used under Android's app
  restrictions and which approaches appear novel relative to prior work.
\end{itemize}

\section{Background and Motivation}
\label{sec:background-motivation}

\textbf{Android kernel branches and patch propagation.}
Android kernels derive from upstream Linux Long-Term Support (LTS) releases.
Google merges these LTS releases with Android-specific patches to produce
Android Common Kernels (ACKs)~\cite{androidKernelOverview,androidCommonKernels}.
Under the Generic Kernel Image (GKI) architecture~\cite{androidGKI}, Google builds
a standardized core kernel binary directly from ACK across all devices and
device manufacturers package hardware-specific drivers into isolated vendor modules.

Security patches propagate through these stages slowly. An upstream fix from Linux mainline must be 
backported to LTS branches, merged into ACK tracking branches, and integrated
into vendor builds before shipping to devices in a firmware update. Prior studies
show that patches often take months or years to reach devices~\cite{zhang2021androidpatch,li2024investigation}.
Because public Git commits disclose code changes before fixes reach downstream
branches or appear in monthly Android Security Bulletins, this delay creates an
\emph{n-day exposure window}: an interval where an upstream fix is public, but
downstream devices remain unpatched and vulnerable~\cite{maar2025doom,li2026they}.

\PP{Autonomous exploitation during the patch gap.}
This propagation delay means that an attacker can continuously monitor
upstream security fixes, and potentially weaponize them against Android devices,
before patches arrive for Android.
This has become increasingly possible with frontier Large Language Models (LLMs) bearing advanced offensive capabilities, as demonstrated by ExploitGym~\cite{wang2026exploitgymaiagentsturn}.
There, we see frontier reasoning models can autonomously generate functional kernel exploits and achieve privilege escalation on Linux targets.

How much these results extend to production Android systems, however, is
unclear. Existing evaluations largely consider Linux virtual machines
running command-line binaries inside an \code{nsjail} sandbox, often provide
the agent with pre-constructed proof-of-vulnerability (PoV) reproducers, and
disable mitigations such as KASLR. Android removes many of these
simplifications. Starting only from a public patch, an agent must determine
whether the vulnerability applies to a particular Android kernel, reconstruct a
trigger without a pre-built reproducer, and develop an exploit that runs from
an unprivileged APK in the \code{untrusted_app} SELinux domain. The exploit
must additionally operate under Android's seccomp restrictions and active
production mitigations, including KASLR, PAN, PXN, and CFI, ultimately obtaining
root privileges and disabling SELinux.

Our goal is therefore not to assume that such exploitation is feasible, but to measure it empirically. This is important not only for understanding how broadly n-day vulnerabilities can be exploited in practice, but also for determining what it would take to continuously manufacture working n-day exploits at scale --- and how effective such a capability is today using publicly accessible LLMs.

\section{Methodology}
\label{sec:methodology}

We design an end-to-end exploit-generation pipeline to evaluate the feasibility of automated Android n-day exploitation across a range of Android Common Kernel and Pixel kernel targets. In this section, we first present an overview of the pipeline, and then describe each stage in detail, including its design rationale and role in the overall workflow.

\vspace{0.02in}
\noindent\textbf{Pipeline Overview.}
Our goal is to efficiently identify upstream bugs that are both relevant to Android and potentially exploitable by ordinary applications. A key challenge is the large number of upstream patches. In principle, we can leverage LLMs to analyze every single upstream patch commit. However, with an average of over 200 commits per day in the upstream~\cite{upstream_commit_count_per_day}, this can be a costly and inefficient continuous operation. To keep the analysis scalable and cost-effective, we design a multi-stage filtering pipeline that progressively narrows the candidate set.

\begin{figure*}[t]
  \centering
  \includegraphics[width=\textwidth]{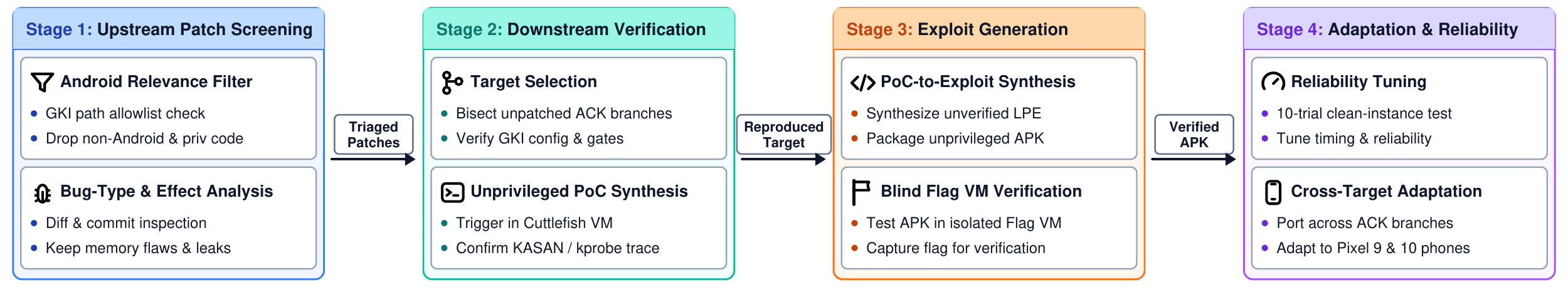}
  \caption{End-to-end vulnerability triage, downstream verification, exploit generation, and cross-target adaptation pipeline.}
  \label{fig:pipeline}
\end{figure*}

As shown in Figure~\ref{fig:pipeline}, our pipeline operates across four stages:
(1)~\textbf{upstream patch screening} applies an allowlist filter and bug-effect analysis to identify memory-corruption flaws;
(2)~\textbf{downstream verification} identifies vulnerable ACK branches via patch bisection, verifies kernel configuration and unprivileged reachability through target selection checks, and synthesizes an unprivileged PoC in Cuttlefish;
(3)~\textbf{exploit generation} advances the PoC to an unverified exploit and validates the final APK under an isolated Flag VM; and
(4)~\textbf{adaptation and reliability} optimizes fragile exploits on Cuttlefish and adapts verified exploits across kernel branches, newer commits, and physical smartphones.
This design reserves expensive synthesis and execution stages for confirmed candidates while systematically evaluating exploit transferability.

\subsection{Upstream Patch Screening}
\label{sec:candidate-triage}

For each upstream patch, we collect the bug-fixing commit (including commit title, commit message, and code diff), the bug-inducing commit (\texttt{fixes: tag}), the pre-fix and post-fix source, and relevant Git history.
To manage the cost of the study, we also have an operating mode limiting the upstream patches to those that are already assigned CVEs, which also provides some additional information useful for filtering, e.g., the bug description in the CVE record.

\PP{Android relevance filter}
We first remove patches outside the Android kernel surface available to ordinary applications. For each upstream patch, the filter uses the affected source paths, and optionally the description from the public CVE record (when the CVE mode is chosen). A patch passes only if at least one affected file matches an allowlist of kernel source paths within the Generic Kernel Image (GKI) (Section~\ref{sec:background-motivation}) which covers subsystems that are reachable by an ordinary application on Android.
The allowlist is conservative by design --- it only excludes files that are certainly not accessible to ordinary apps, e.g., raw sockets and debugfs. 
If the CVE description is provided, there are also blocklist rules that reject the bug if its description explicitly states it requires elevated capabilities, e.g., requiring \texttt{CAP\_SYS\_ADMIN}. The blocklist is also a conservative filter.
The Android relevance filter is cheap as it does not involve LLMs.

\PP{Bug effect filter}
For each upstream patch that passes the Android relevance filter, an LLM reviews the
bug-fixing commit, pre-fix implementation, surrounding code, and Git history to
determine whether the patch fixes a bug that leads to memory corruption or information
disclosure. We conservatively keep bugs that may cause either effect,
including inconclusive cases, and reject those whose effect is limited to issues such as resource leaks or denial-of-service.

\subsection{Downstream Verification}
\label{sec:runtime-reproduction}
\label{sec:patch-propagation}
\label{sec:propagation-method}

Upstream screening identifies bugs with memory-corruption potential, but does
not determine whether the defect exists in downstream Android releases or
whether an unprivileged application can trigger it at runtime. Downstream
verification addresses this across three steps:
(1)~we identify which Android Common Kernel (ACK) branches contain the
defect using patch bisection across ACK branches;
(2)~we apply Android target selection checks to verify that candidate targets
contain the vulnerable code, enable required kernel configurations, and
expose the interface to unprivileged applications; and
(3)~we run the target in a Cuttlefish virtual machine and synthesize
an unprivileged reproducer that triggers the defect.

\PP{Branch-Specific Vulnerable Window Determination}
\label{sec:vulnerable-window-determination}
To identify which downstream releases are vulnerable and when patches land,
we perform commit tracking across upstream (Linux) and downstream (ACK). 
Following the methodology of patch-tracking from Zhang
et~al.~\cite{zhang2021androidpatch}, we track when an upstream patch is backported to downstream, using commit subjects, upstream references in commit messages, and source change sites.

For each ACK branch, we define the
\emph{patch boundary} as the earliest commit along the branch's mainline (first-parent) history that contains the fix. If a later patch repairs that
initial fix (e.g., due to the initial one being an incomplete fix), we advance the patch boundary to that subsequent commit. When a
patch lands on another branch and merges into an ACK branch later, the branch
incorporates the fix at that merge commit.

Unlike Zhang et~al., who track only patch commits to measure propagation delay,
we also identify when each vulnerability was introduced using a similar method. We locate the bug-inducing commit using the upstream patch's
\code{Fixes:} trailers. We then trace these bug-inducing commits and their equivalent downstream history to determine when the vulnerability first appears on each branch, which we refer to as a \emph{bug-introduction boundary}. When multiple bug-inducing commits are listed, we conservatively pick the last of those commits as the boundary, to not overclaim the vulnerable window.

We analyze each CVE--branch pair for these boundaries. When both boundaries are
known, the vulnerable window of the branch begins with the introduction commit
and ends immediately before the patch boundary. We call this a 
\emph{closed range}. If no patch boundary appears, the range remains open through the branch head. When the introduction and patch appear
in the same commit, the branch was never vulnerable. Pairs where we cannot identify
the introduction or patch commit remain undetermined.

\PP{Android target selection checks}
Using the vulnerable branch ranges established earlier, this step
selects a concrete target release and verifies its eligibility before runtime
reproduction. This step is necessary because the exact target configuration and downstream kernel may have behaviors that are not captured by boundary mapping.
For each candidate bug, a target-selection LLM receives the
upstream fixing commit, the affected source files, and the available ACK
branches and release tags. It selects a branch and release tag dated within the
12 months before the upstream fix. We call this branch and release tag a
\emph{candidate target}.

The LLM analyzes the candidate target and checks whether:

\begin{enumerate}[nosep,leftmargin=*]
  \item the vulnerable logic is present and neither the upstream fix nor an
        equivalent correction is present;
  \item the required kernel configuration and interfaces are available;
  \item the affected path is not permission- or privilege-gated; and
  \item the bug still has the expected memory-corruption or
        information-disclosure effect.
\end{enumerate}

If any check fails, the LLM selects a different eligible branch or release tag
and repeats the analysis. This stage repeats at most three times per bug and
stops when a target passes all four checks. The passing target proceeds to
runtime reproduction. If no target passes after three attempts, the bug does
not proceed.

\PP{Runtime reproduction in Cuttlefish}
For each bug that passes the Android target selection checks, we build the
selected kernel using \code{gki_defconfig} (with KASAN enabled to aid bug detection) and boot it in a Cuttlefish VM. 
We use \emph{target} to mean
this specific ACK release, built kernel, and Cuttlefish configuration.

A reproduction agent, separate from the target-selection LLM, performs this
stage in the same isolated, sandboxed, and network-restricted container
environment described in Section~\ref{sec:agent-environment}.

The reproduction agent develops and tests a reproducer iteratively on the
selected target. It can revise its test case, inspect kernel logs, use kprobes
to observe the suspected faulty state, and repeat executions based on the
results. Evidence may take the form of a Kernel Address Sanitizer (KASAN)
report, a kprobe observation, or a kernel crash.
The reproduction prompt is straightforward: it provides the upstream commit title, message, and code diff, explains the interface for interacting with the VM, and instructs the agent to write a reproducer that triggers the bug from an unprivileged app with no other guidance.

Runtime reproduction permits at most three target attempts per bug. The first
attempt uses the target that passed the Android target selection checks. If
reproduction fails, the target-selection LLM chooses another branch or release
tag from the same 12-month window; we build and boot the new target, and the
reproduction agent tries again. The stage stops when an attempt meets the
criteria described below. After three failed runtime attempts, the bug is a
reproduction failure. We discard failed attempts and pass successful
CVE--target pairs forward through the pipeline.

\PP{Evidence and success criteria}
\label{sec:discovery-success}
A successful reproduction must satisfy both a trigger requirement and an
evidence requirement. The initiating test case must run from an installed
Android application with an ordinary application UID in the
\code{untrusted_app} SELinux domain. Native code qualifies when the
application launches it. A native binary started from \code{adb shell},
\code{su}, or another privileged domain does not qualify.

The runtime evidence must connect the observed effect to the vulnerability
repaired by the upstream patch. We accept a root-cause-specific KASAN report, a
kernel crash with a causal trace through the affected code, confirmed
disclosure of kernel data, or a witness of memory corruption such as bad
page-cache state, silent data corruption (confirmed by kprobe traces). When these
methods do not expose the fault, kprobes may be used to record the bug-specific
faulty state as evidence. 

\subsection{Exploit Generation}
\label{sec:exploit-generation}
\label{sec:exploit-task}

The exploit-generation evaluation uses the CVE--target pairs that satisfy
the discovery and reproduction criteria. For each pair, we resume the LLM
session used for reproduction in the same environment and send a new message
asking the agent to develop an exploit. This message changes the objective but
provides no additional CVE or target information, and it does not prescribe
exploitation techniques or templates. Because we resume the
conversation, the agent maintains its source analysis, files, reproducer, runtime evidence,
and conversation context from reproduction (Section~\ref{sec:runtime-reproduction}). The exploit prompt simply instructs the agent to turn the reproduced bug into a local privilege escalation to root, set SELinux to permissive, and package the attack as an unprivileged Android application that retrieves a protected flag file across reboots, providing no other guidance.

This task models an attacker who knows the public upstream fix and has
identified an affected Android kernel release. The final attack begins in an
installed application running in the \code{untrusted_app} SELinux domain, under the standard seccomp profile, with production mitigations active, and
without root or administrative privilege. The development VM
provides root access for debugging and observation, but the final application
cannot depend on a privileged bootstrap or instrumentation to trigger the
vulnerability or complete the privilege escalation.

The LLM submits its application to a blind evaluator that tests whether it
obtains root, changes SELinux to permissive, and reads a protected secret.

\PP{Isolated long-horizon agent environment}
\label{sec:agent-environment}
Reproduction and exploit agent work takes place in a persistent Docker
container dedicated to one CVE--target pair. The container provides the
upstream fix, source for the selected target, the Android Native Development
Kit (NDK), a private workspace, and root access to one assigned Cuttlefish VM
or, when required, a real phone.
Workspace and agent state persist across reproduction, exploit development,
and APK packaging.

Each experiment is subject to a fixed model-cost budget ceiling. A
host-side monitor tracks the cumulative cost from the session's token counts
and stops the experiment when it reaches this ceiling. For consistency, 
we use the per-token pricing 
published at the beginning of the evaluation period for all experiments (detailed in Section~\ref{sec:evaluation}).

The container runs with Docker \code{--network none} and has no direct route
to the host, local network, or public Internet. Network access is limited to
hostnames required for the model service and package repositories for Ubuntu,
Debian, and Python. Hosted web search is disabled and internet access is
restricted so that the agent cannot copy exploits that are already made public. The agent can still install ordinary development
dependencies from the allowed package repositories.

Each experiment can access only its assigned Cuttlefish instance or real phone
through an ADB relay and a narrow controller interface. Separate interfaces
provide brokered network access and blind evaluation. With this sandbox, agents
cannot access the host, other experiments' workspaces, unassigned Cuttlefish
VMs or phones, or retrieve information from the general Web, such as existing
exploit implementations.

\phantomsection
\label{sec:cve-attribution}
\PP{Blind verification and CVE attribution}
We do not accept the LLM agent's claim of success without independent
verification. After local testing, the agent submits its APK to an automated, deterministic (non-LLM) evaluator that runs outside the experiment environment. The evaluator provisions a separate Cuttlefish VM with the target kernel and Android image. The agent
cannot access this VM through ADB, a shell, logs, or the network.

The evaluator places a fresh secret in a root-protected file, installs the APK,
and launches it. A successful APK must start as an ordinary Android
application, obtain root, change SELinux to permissive, and read the secret.
The application must then write the secret to a file in its private data
directory.
The evaluator checks these conditions from outside the VM and returns only a
verdict and a bounded failure reason. A failure can be investigated on the
development VM before the agent submits a revised APK.

For each passing submission, the evaluator archives the exact APK and its
tracked source and creates a receipt containing the artifact digest. This binds
the verified result to the submitted application and preserves the artifact
used for later analysis and replay.

Finally, a blind pass establishes privilege escalation but does not identify the
vulnerability responsible for it. ExploitGym~\cite{wang2026exploitgymaiagentsturn}
demonstrated that flag capture alone can be satisfied by unrelated software bugs
or environmental shortcuts, and introduced an LLM agent-as-a-judge to evaluate
whether an exploit genuinely exercised the intended vulnerability. We adopt
this approach and ask a separate LLM reviewer to audit every passing artifact.
The reviewer first identifies the defect repaired by the upstream fix, then
traces the archived exploit from its trigger through the first useful primitive
and the final privilege escalation. It checks the archived source, development
history, runtime evidence, session transcript, evaluator receipt, and replay
results for evidence of an unrelated vulnerability, privileged assistance, or
stale evaluator state. If the reviewer confirms that the exploit uses the
assigned vulnerability, the artifact passes verification; otherwise, it is a
failure.

\subsection{Reliability and Adaptation}
\label{sec:adaptation-reliability}

After an exploit passes initial verification on Cuttlefish, the final stage of
our pipeline evaluates exploit reliability and cross-target adaptation. We
first evaluate the exploit's execution reliability across clean boots on
Cuttlefish and improve it when necessary. We then adapt the exploit across
software evolution (newer commits and adjacent kernel branches) and to physical
smartphones. This is to measure how much additional effort is required to transfer it across related Android targets, including physical phones. 

\PP{Reliability evaluation and optimization on Cuttlefish}
To evaluate exploit reliability against runtime nondeterminism (such as race
timing and memory layout randomization), we execute each verified application
across repeated, clean-instance trials on the original target under the blind
evaluator. We record whether each trial successfully obtains root and reads the
secret without crashing the kernel.  
When an exploit exhibits low reliability (falling below 80\% across repeated trials), the pipeline prompts the agent to improve the reliability (again, with no other guidance). The agent then revises its timing, heap conditioning, or retry logic to improve consistency across boots.

\PP{Adaptation across kernel branches and commits}
To evaluate exploit transferability across kernel evolution and prepare for future physical phone releases based on those
phone versions, we prompt the
agent to adapt verified exploits across newer commits on the same branch and to
adjacent ACK branches. In these experiments, the agent receives the updated
target kernel source and the same development environment, and must modify
its exploit primitives, structure offsets, and timing to achieve a blind pass on
the new target.

\PP{Physical device adaptation}
Using the unpatched factory builds identified by our vulnerable window
determination (Section~\ref{sec:vulnerable-window-determination}), we flash
Google Pixel smartphones to evaluate real-hardware adaptation. During this
stage, the agent retargets the verified Cuttlefish exploit to the phone's
production kernel and runtime environment. An automated, deterministic on-device evaluator then tests whether the adapted exploit independently achieves local privilege escalation and reads the protected secret from the device.

\section{Pipeline Results}
\label{sec:evaluation}
\label{sec:exploit-results}

This section evaluates empirical outcomes across our pipeline. We first report the triage funnel and runtime reproduction results, followed by unverified and verified exploit milestones, agent time, and model costs of autonomous exploit generation. We then report exploit reliability and adaptation results on virtual and physical devices.

\PP{Experimental setup}
To manage the cost of the experiments, we limit our pipeline to the upstream patches that are already associated with Linux kernel CVEs. We choose those that are disclosed between January 2024 and September 2026. This leads to a total of 10,301 CVEs and upstream patches (2,831 in 2024, 2,439 in 2025, and 5,031 in 2026).
Experiments run using OpenAI's \code{gpt-5.6-sol} reasoning model with \code{xhigh} reasoning effort under a \$1,250 per-experiment cost ceiling. 
We estimate model cost from recorded token usage using the published pricing at the time of experiment: \$5.00 per million uncached input tokens, \$0.50 per million
cached input tokens, \$6.25 per million cache-write input tokens, and \$30.00 per
million output tokens.

Cuttlefish virtual machine environments span Android 14 through Android 17 images, each configured with six vCPUs. Physical validation runs on commercial Google Pixel~10 (Android 16) and Pixel~9 (Android 16 and 14) smartphones reflashed with unpatched factory builds. We parallelize experiments when convenient on an \code{aarch64} server with 128 CPUs at 3.0\,GHz, 1\,TB RAM, and 3.5\,TB storage.

\subsection{Upstream Patch Screening Outcomes}

Across 10,301 evaluated upstream CVEs, the automated triage funnel retains 1,233
vulnerabilities relevant to Android --- a 12.0\% pass rate
(Table~\ref{tab:android-filter-by-year}). Subsequent filtering and target
selection narrow to 38 reproducible targets (Table~\ref{tab:discovery-funnel}).
This lower applicability rate is unsurprising: Android kernels enable only a subset of upstream Linux functionality, and Android's SELinux, seccomp, and application sandbox further restrict the kernel interfaces reachable from untrusted applications.

\begin{table}[t]
  \centering
  \small
  \caption{Android relevance filter results by CVE year.
    }
  \label{tab:android-filter-by-year}
  \begin{tabular}{lrr}
    \toprule
    \textbf{CVE year} & \textbf{Raw CVEs} & \textbf{Passed relevance filter} \\
    \midrule
    2024 & 2,831 & 388 \\
    2025 & 2,439 & 301 \\
    2026 & 5,031 & 544 \\
    \midrule
    \textbf{Total} & \textbf{10,301} & \textbf{1,233} \\
    \bottomrule
  \end{tabular}
\end{table}

The bug effect filter retains 612 of the 1,233 relevance-filtered candidates
(49.6\%; Table~\ref{tab:discovery-funnel}) at an average model cost of \$0.61
per candidate (\$748.74 across all 1,233 reviews). 

Considering that our analysis covers all CVEs reported over a 33-month period (January 2024 to September 2026), the screening cost is relatively low. In contrast, exhaustively screening upstream commits would be substantially more expensive. We identified over 200,000 upstream commits during the same period. Assuming a comparable relevance-filtering rate, we estimate that the cost would increase by approximately 20-fold to \$14,960. Moreover, this estimate accounts only for the first stage of the pipeline, excluding the costs of subsequent analysis and exploit generation, which can be even more expensive.

\subsection{Downstream Verification Outcomes}

\begin{table}[t]
  \centering
  \small
  \caption{CVE filtering and reproduction funnel.}
  \label{tab:discovery-funnel}
  \begin{tabularx}{\columnwidth}{@{}Xrr@{}}
    \toprule
    \textbf{Stage outcome} & \textbf{CVEs} & \textbf{Prior-stage share} \\
    \midrule
    Collected CVEs & 10,301 & --- \\
    Passed Android relevance filter & 1,233 & 12.0\% \\
    Passed bug effect filter & 612 & 49.6\% \\
    Passed Android target selection checks & 152 & 24.8\% \\
    Satisfied reproduction criteria & 38 & 25.0\% \\
    \bottomrule
  \end{tabularx}
\end{table}

From the 612 bugs that are likely to affect Android, the pipeline performs vulnerable-window determination in Android Common Kernel branches, and target-selection checks then identify 152 bugs affecting Cuttlefish targets (Table~\ref{tab:discovery-funnel}). Target-selection reviews consumed \$415.51 in model costs across the 612 CVEs (median \$0.59, mean \$0.68).

Out of the 152 bugs, the pipeline successfully reproduced 38 bugs on 18 selected targets with kernel versions ranging from 5.10.223 to 6.18.21.
Across all 152 attempted CVEs, runtime reproduction consumed \$1,128.56 in model costs (median \$4.02, mean \$7.42).
The 38 successful reproductions accounted for \$702.50 (median \$8.41, mean \$18.49; Table~\ref{tab:reproduction-costs}).
The 114 failed attempts consumed \$426.05 (median \$2.70, mean \$3.74), costing substantially less per vulnerability because the agent exited early when encountering blockers: over 60\% were blocked by unprivileged permission restrictions, roughly 20\% by disabled kernel configurations, and the remainder by other target environment constraints. 

We see an interesting correlation between the reproduction cost and successful exploit generation subsequently. For the 17 targets that produced a verified exploit later on, reproduction cost totaled \$154.39 (median \$6.45).
For the remaining 21 that failed exploit generation, reproduction costs were higher on average (\$548.12 across 21 targets; mean \$26.10).
This is likely because the agent spent more turns navigating complex
bugs, e.g., race conditions, that are both difficult to trigger and exploit.

\begin{table}[t]
  \centering
  \footnotesize
  \setlength{\tabcolsep}{2pt}
  \caption{Model reproduction costs for all 38 reproducible CVE--target pairs. Asterisks ($^*$) indicate the 17 targets that produced a verified LPE. }
  \label{tab:reproduction-costs}
  \begin{tabular}{@{}llr@{\hspace{6pt}}llr@{}}
    \toprule
    \textbf{CVE} & \textbf{Target} & \textbf{Cost} & \textbf{CVE} & \textbf{Target} & \textbf{Cost} \\
    \midrule
    CVE-2024-26640$^*$ & 6.1.68   & \$4.25  & CVE-2026-23161     & 6.12.58  & \$5.11  \\
    CVE-2024-26759     & 6.1.25   & \$7.73  & CVE-2026-23394     & 6.1.145  & \$6.43  \\
    CVE-2024-26923     & 6.1.25   & \$18.06 & CVE-2026-31397     & 6.18.12  & \$3.97  \\
    CVE-2024-36904     & 6.1.78   & \$50.26 & CVE-2026-31398     & 6.18.12  & \$14.21 \\
    CVE-2024-36971$^*$ & 6.1.68   & \$11.36 & CVE-2026-31648$^*$ & 6.18.12  & \$10.52 \\
    CVE-2024-36972$^*$ & 6.1.78   & \$6.45  & CVE-2026-43074     & 6.18.12  & \$14.43 \\
    CVE-2024-38580$^*$ & 6.1.68   & \$6.37  & CVE-2026-43284$^*$ & 6.18.12  & \$5.10  \\
    CVE-2024-44987     & 6.1.25   & \$32.20 & CVE-2026-43499$^*$ & 6.18.0   & \$7.23  \\
    CVE-2024-46740$^*$ & 6.1.25   & \$6.24  & CVE-2026-46215$^*$ & 6.18.12  & \$5.67  \\
    CVE-2024-53079     & 6.6.30   & \$59.83 & CVE-2026-46242     & 6.18.12  & \$35.82 \\
    CVE-2024-56554$^*$ & 6.6.50   & \$4.94  & CVE-2026-52910     & 6.1.57   & \$2.94  \\
    CVE-2024-56555     & 5.10.223 & \$74.65 & CVE-2026-52973$^*$ & 6.18.12  & \$6.42  \\
    CVE-2025-38349     & 6.12.23  & \$12.88 & CVE-2026-53361$^*$ & 6.18.0   & \$9.27  \\
    CVE-2025-38447     & 6.6.127  & \$12.39 & CVE-2026-53362$^*$ & 6.18.12  & \$5.42  \\
    CVE-2025-39977$^*$ & 6.12.38  & \$6.87  & CVE-2026-53366     & 6.18.12  & \$4.81  \\
    CVE-2025-40135     & 5.15.110 & \$78.69 & CVE-2026-63816     & 6.18.12  & \$89.57 \\
    CVE-2025-40158     & 5.15.74  & \$16.32 & CVE-2026-63952     & 6.18.12  & \$3.37  \\
    CVE-2025-40201     & 6.1.57   & \$4.45  & CVE-2026-64560$^*$ & 6.1.162  & \$10.73 \\
    CVE-2026-23077$^*$ & 6.18.0   & \$9.09  & CVE-2026-80778$^*$ & 6.18.21  & \$38.44 \\
    \midrule
    \multicolumn{3}{@{}l}{\textbf{Total (38 targets)}} & \multicolumn{3}{r@{}}{\textbf{\$702.50 (median \$8.41)}} \\
    \bottomrule
  \end{tabular}
\end{table}

\PP{Cost comparison}
To evaluate whether our staged CVE filtering is necessary, we compare it
against a naive baseline that assesses upstream CVEs directly against an ACK
release without filtering or vulnerable window determination using commit tracing. In this baseline, an agent
receives read-only access to local ACK Git history, the CVE record, and the
upstream patch. The agent evaluates whether the target commit retains the vulnerable code, whether the bug
causes memory corruption, whether Android GKI configuration makes the code
reachable, and whether Android runtime or kernel permission checks gate the
trigger.

Across a random sample of 25 CVEs, 18 (72.0\%) retained the vulnerable code and
17 (68.0\%) involved memory corruption. Requiring all reachability and
unprivileged-trigger conditions identified 2 of 25 CVEs (8.0\%) as viable PoC
candidates. The average model cost per review was \$0.93. Extrapolating this
cost across 10,000 upstream CVEs would exceed \$9,200 for a single downstream
target, requiring repeated expenditure for each additional kernel release.
In contrast, our screening cost applies globally across all targets: the bug-effect
filter costs \$748.74 in total, vulnerable-window determination identifies affected releases across all ACK branches at negligible cost, and target-selection checks add \$415.51 across the 612 candidates (averaging \$0.68 per CVE).
Operating the entire triage pipeline across all targets thus costs \$1,164.25, compared to over \$9,200 for a single target under the naive approach.

\subsection{Autonomous Exploit-Generation Outcomes}
\label{sec:exploit-outcomes}

Table~\ref{tab:exploit-results} summarizes the furthest milestone reached by the 38 CVE--target pairs that enter exploit generation. Across the full pipeline, 17 of the 152 CVEs (11.2\%) that pass Android target selection and enter runtime reproduction ultimately yield a verified LPE. These successful exploits span nine recorded target kernel versions, from 6.1.25 through 6.18.21. Our CVE-attribution review (Section~\ref{sec:cve-attribution}) confirms that all 17 verified APKs exploit their assigned vulnerabilities. Only two runs that progress beyond exploit generation subsequently fail during packaging or blind verification. By contrast, half of the 38 evaluated CVE--target pairs make no further progress after runtime reproduction.

\begin{table}[!htbp]
  \centering
  \small
  \caption{Exploitation outcomes and terminal reasons across all 38 evaluated CVE--target pairs.}
  \label{tab:exploit-results}
  \begin{tabularx}{\columnwidth}{@{}Xrr@{}}
    \toprule
    \textbf{Outcome / Terminal reason} & \textbf{Cases} & \textbf{Share} \\
    \midrule
    \textbf{Verified LPE} & \textbf{17} & \textbf{44.7\%} \\
    \midrule
    \textbf{Unverified LPE} & \textbf{2} & \textbf{5.3\%} \\
    \quad Budget exhausted during blind verification & 2 & 5.3\% \\
    \midrule
    \textbf{Reproduction only} & \textbf{19} & \textbf{50.0\%} \\
    \quad Budget exhausted before initial LPE & 14 & 36.8\% \\
    \quad Agent concluded no viable LPE path & 5 & 13.2\% \\
    \midrule
    \textbf{Total} & \textbf{38} & \textbf{100.0\%} \\
    \bottomrule
  \end{tabularx}
\end{table}

\PP{Failure analysis.} Table~\ref{tab:exploit-results} also classifies why 21 experiments do not
produce a verified LPE, based on session logs and cost-shutoff records. Most
failures stem from budget exhaustion rather than an inability to progress
beyond the initial reproduction.

In most budget-exhausted runs, the agent
constructs intermediate primitives---such as race orchestration, heap
conditioning, or probe-assisted memory corruption---but exhausts the cost
ceiling while attempting to eliminate development scaffolding (e.g., kprobe
timing holds or per-boot pointer disclosures) to create a standalone,
unprivileged APK. 
In two cases, the agent produced a working local LPE during
development but exhausted the budget while attempting to stabilize execution
across clean boots during blind verification. 
In the five runs terminated by the
agent, we find that the agent identified constraints that prevented
an LPE path, such as out-of-bounds corruption confined to packet tailroom (i.e., padding),
read-only mappings lacking backing page descriptors, or the absence of a
reachable privileged consumer. 

\PP{Cross-validate with published exploits.}
To evaluate exploit generation against human ground truth, we audited the 38 evaluated CVEs for pre-existing public privilege-exploit artifacts (e.g., from kernelCTF~\cite{kernelctf} or public security write-ups) that predated our experiment runs.
Ten of the 38 CVEs had such published privilege exploits.
Because the evaluation environment disables general Internet and hosted web search, the agent cannot retrieve or copy these public artifacts.
Across these ten CVEs, the pipeline synthesized a verified LPE for seven (70.0\%):
CVE-2024-36972~\cite{ret2bpf},
CVE-2026-43284~\cite{dirtyfrag},
CVE-2026-43499~\cite{ghostlock},
CVE-2026-46215~\cite{cve202646215},
CVE-2026-53361~\cite{badgarbage},
CVE-2026-53362~\cite{ipv6fragescape}, and
CVE-2026-64560~\cite{cybermeowfia64560}.

The remaining three CVEs with published exploits (CVE-2026-23394~\cite{cve202623394kernelctf}, CVE-2026-43074~\cite{cybermeowfia43074}, and CVE-2026-46242 (Bad Epoll)~\cite{badepoll}) all terminated due to budget exhaustion.
For CVE-2026-23394, the agent encountered a race-engineering failure: while a shell-launched debug exploit reached permissive SELinux using a kernel probe to pause the socket reader, attempts to widen the race from an unprivileged app proved too fragile, either missing the race window or crashing the kernel.
For CVE-2026-43074, the agent suffered a similar failure: while a standalone ADB-shell program attained root and permissive SELinux, converting it into an APK altered process heap layout and descriptor state, preventing a successful exploit from an unprivileged app.
For CVE-2026-46242, an unprivileged APK reached root and disabled SELinux locally (yielding an unverified LPE), but collateral kernel corruption stalled subsequent system calls, and the agent exhausted its budget attempting to complete the flag retrieval for blind verification.
We discuss these failure modes in Section~\ref{sec:discussion}.

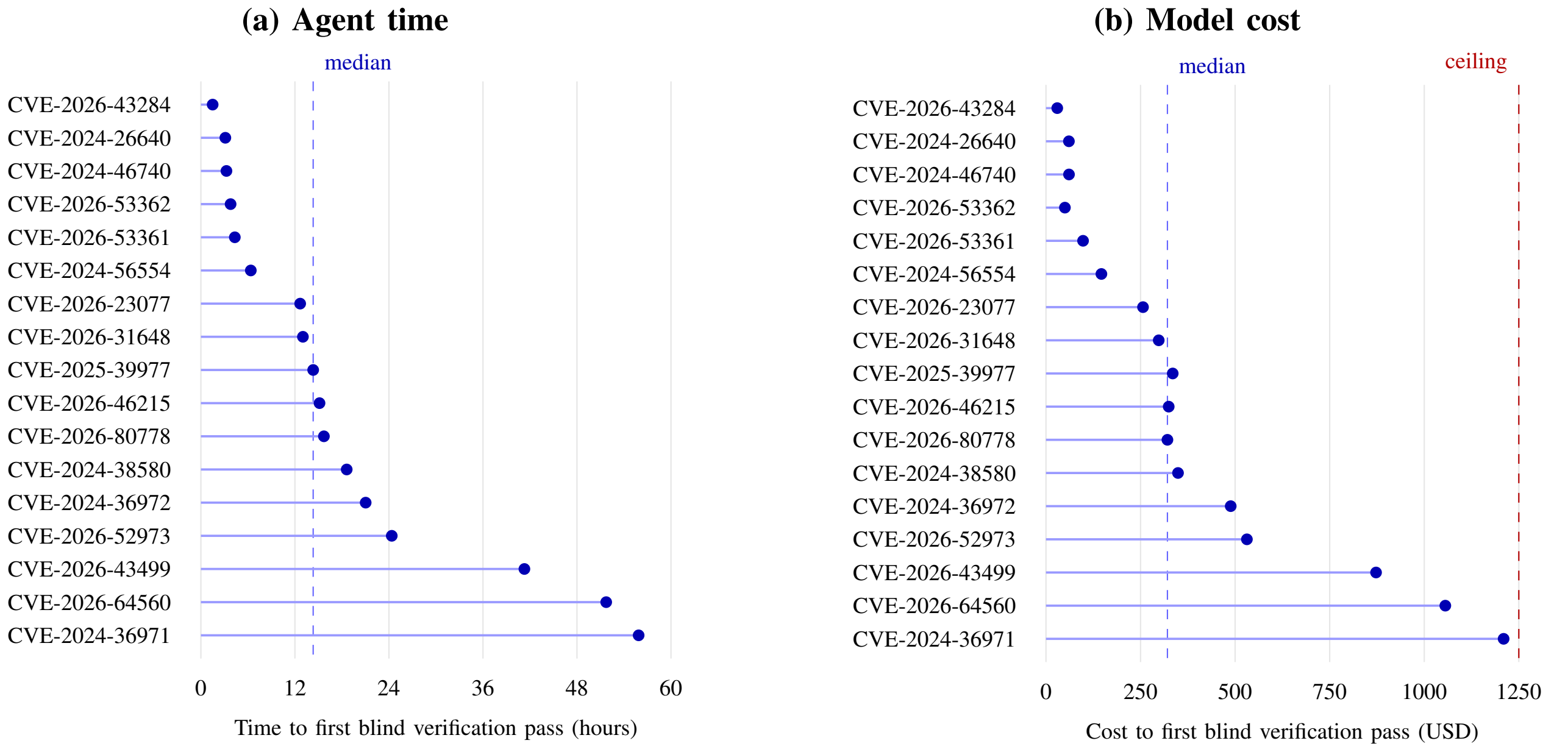
\begin{figure*}[!t]
  \centering
  \begin{minipage}[t]{0.48\textwidth}
    \centering
    \textbf{(a) Agent time}

    \vspace{2pt}
    \begin{tikzpicture}[x=0.085cm,y=0.36cm,font=\scriptsize]
      \foreach \hours in {0,12,24,36,48,60} {
        \draw[gray!20] (\hours,0.7) -- (\hours,-16.7);
        \node[anchor=north] at (\hours,-17.0) {\hours};
      }
      \draw[dashed,blue!60] (14.33,0.7) -- (14.33,-16.7);

      \foreach \name/\hours [count=\i from 0] in {
        CVE-2026-43284/1.50,
        CVE-2024-26640/3.12,
        CVE-2024-46740/3.26,
        CVE-2026-53362/3.79,
        CVE-2026-53361/4.33,
        CVE-2024-56554/6.37,
        CVE-2026-23077/12.67,
        CVE-2026-31648/13.02,
        CVE-2025-39977/14.33,
        CVE-2026-46215/15.14,
        CVE-2026-80778/15.70,
        CVE-2024-38580/18.61,
        CVE-2024-36972/21.03,
        CVE-2026-52973/24.36,
        CVE-2026-43499/41.29,
        CVE-2026-64560/51.73,
        CVE-2024-36971/55.87
      } {
        \pgfmathsetmacro{\yy}{-\i}
        \node[anchor=east,xshift=-2mm] at (0,\yy) {\name};
        \draw[blue!40,line width=0.7pt] (0,\yy) -- (\hours,\yy);
        \fill[blue!70!black] (\hours,\yy) circle (1.8pt);
      }

      \node[anchor=south west,blue!70!black] at (14.33,0.7) {median};
      \node[anchor=north] at (30,-18.2)
        {Time to first blind verification pass (hours)};
    \end{tikzpicture}
  \end{minipage}
  \hfill
  \begin{minipage}[t]{0.48\textwidth}
    \centering
    \textbf{(b) Model cost}

    \vspace{2pt}
    \begin{tikzpicture}[x=0.0041cm,y=0.36cm,font=\scriptsize]
      \foreach \cost in {0,250,500,750,1000,1250} {
        \draw[gray!20] (\cost,0.7) -- (\cost,-16.7);
        \node[anchor=north] at (\cost,-17.0) {\cost};
      }
      \draw[dashed,blue!60] (321.07,0.7) -- (321.07,-16.7);
      \draw[dashed,red!70!black] (1250,0.7) -- (1250,-16.7);

      \foreach \name/\cost [count=\i from 0] in {
        CVE-2026-43284/29.80,
        CVE-2024-26640/60.64,
        CVE-2024-46740/60.85,
        CVE-2026-53362/50.05,
        CVE-2026-53361/98.01,
        CVE-2024-56554/146.50,
        CVE-2026-23077/256.51,
        CVE-2026-31648/298.08,
        CVE-2025-39977/335.05,
        CVE-2026-46215/324.45,
        CVE-2026-80778/321.07,
        CVE-2024-38580/349.03,
        CVE-2024-36972/488.36,
        CVE-2026-52973/531.17,
        CVE-2026-43499/872.57,
        CVE-2026-64560/1055.70,
        CVE-2024-36971/1209.82
      } {
        \pgfmathsetmacro{\yy}{-\i}
        \node[anchor=east,xshift=-2mm] at (0,\yy) {\name};
        \draw[blue!40,line width=0.7pt] (0,\yy) -- (\cost,\yy);
        \fill[blue!70!black] (\cost,\yy) circle (1.8pt);
      }

      \node[anchor=south west,blue!70!black] at (321.07,0.7) {median};
      \node[anchor=south east,red!70!black] at (1250,0.7) {ceiling};
      \node[anchor=north] at (625,-18.2)
        {Cost to first blind verification pass (USD)};
    \end{tikzpicture}
  \end{minipage}
  \caption{Agent time and cumulative model cost from the
  reproduction prompt, both measured through
  the first blind verification pass for each of the 17 verified CVE--target pairs. The red
  line marks the \$1,250 cumulative per-experiment cost ceiling.}
  \label{fig:exploit-development-time}
\end{figure*}

\PP{Cost analysis.}
Figure~\ref{fig:exploit-development-time} reports agent development time and cumulative model cost from the initial reproduction prompt through the first blind verification pass. We sum active elapsed time within logged agent turns, including model inference, tool execution, and synchronous waits within a turn. As plotted in Figure~\ref{fig:exploit-development-time}(a), the median time to a verified exploit is 14.3 agent-hours, with most experiments completing within a single day.

Across the 17 verified exploits, the median development cost is \$321.07
(totaling \$6,487.67, with initial reproduction accounting for \$154.39;
Table~\ref{tab:reproduction-costs}; Figure~\ref{fig:exploit-development-time}(b)).
Upstream patch screening cost \$748.74 across candidate CVEs, target selection added \$415.51, and the 114 failed reproduction attempts consumed \$426.05.
Accounting for all 38 full exploit runs from reproduction through completion (\$6,487.67 across the 17 verified exploits and \$20,692.32 across the 21 unsuccessful attempts), the pipeline consumed \$28,770.29, or \$1,692 per verified LPE.
This demonstrates that autonomous n-day exploitation remains economically practical even when factoring in all upstream screening, discarded targets, and budget-exhausted attempts.
The time calculation excludes gaps between agent turns, and neither measurement
includes work after the first pass.

\subsection{Reliability and Adaptation Outcomes}

\PP{Exploit reliability on Cuttlefish}
Repeated trials on clean virtual-machine instances measure whether verified
exploits succeed consistently despite runtime nondeterminism, such as race
timing and address-space randomization. Across 170 completed fresh-instance
trials, i.e., ten per verified CVE--target pair, the overall median reliability is
90\% (Appendix Table~\ref{tab:reliability}). Twelve exploits reached 80\%--100\% reliability
on their initial pass. Five underwent follow-up optimization (as mentioned in \S\ref{sec:adaptation-reliability}) costing \$605.76 in total;
expenditure varied widely, with CVE-2024-36971 accounting for over half (\$305.43)
while remaining at 30\%, whereas CVE-2024-46740 took only \$7.76 to improve from
40\% (4/10) to 100\% (10/10). Overall, 15 of the 17 exploits achieve at least 80\%
reliability (seven at 100\%), and only two exhibit substantial fragility:
CVE-2024-36971 (30\%) and CVE-2026-64560 (10\%).

Exploit reliability depends primarily on two factors: \emph{primitive determinism}
(whether exploitation requires winning a race) and \emph{failure recoverability}
(whether race misses can be retried safely without crashing the kernel). The seven
exploits that achieve 100\% reliability rely on deterministic primitives or safe retries
where pre-execution checks verify memory layout before triggering writes or further
exploit stages. Conversely, fragility stems from race-window sensitivity and asynchronous
kernel activity: background network workers disrupt socket race timing in CVE-2024-36971,
and sub-microsecond race windows or memory compaction disrupt heap conditioning in
CVE-2026-64560.

\PP{Cross-target Cuttlefish adaptation outcomes}
We evaluate whether an agent can port an existing verified exploit when kernel
structures, allocator configurations, or compiler toolchains change across
Cuttlefish targets. Starting from the verified development workspace, the agent
executed 34 paired adaptation experiments covering all 17 verified CVE--target
pairs across two dimensions: same-branch forward-porting (to the newest
vulnerable commit on the same branch) and adjacent-branch cross-porting (to an
adjacent kernel release branch).

All 34 adaptation experiments successfully produced an exploit that passed
independent blind verification on the destination kernel build
(Appendix Table~\ref{tab:adaptation-costs}). Same-branch forward-porting succeeded in all
17 cases at a median model cost of \$11.59 per CVE (totaling \$1,097.49).
Adjacent-branch cross-porting likewise succeeded in all 17 cases at a median
model cost of \$32.51 per CVE (totaling \$3,105.59). Adjacent-branch porting
cost $2.83\times$ more than same-branch porting because differing kernel
versions frequently required the agent to replace missing payload objects or
rebuild gadget chains, whereas same-branch updates require only minor adaptations, e.g., updating constants such as hardcoded offsets and addresses.

\PP{Physical device adaptation outcomes}
Following exploit verification on Cuttlefish, we evaluated adaptation across
nine CVEs on commercial smartphones: Google Pixel~10 (running
Android 16) and Google Pixel~9 (running Android 16 and Android 14). While
Cuttlefish virtual machines can be automatically provisioned and reset via
software controllers, evaluating on hardware requires reflashing each device
with a factory firmware build containing an unpatched kernel. Because reflashing
and device recovery require manual operator intervention, we evaluate
adaptation across a subset of nine verified exploits.

Table~\ref{tab:physical-adaptation} summarizes the physical device adaptation
outcomes across 14 executed trials. In 12 of the 14 trials (85.7\%), the agent
produced an unprivileged exploit that passed blind verification on the device.
All six evaluated CVEs passed on Pixel~10 running Android 16; three of four
passed on Pixel~9 running Android 16; and three of four passed on Pixel~9
running Android 14. Across the nine distinct evaluated CVEs, autonomous
adaptation succeeded on at least one device. In the remaining two
trials, timing and scheduler differences for CVE-2025-39977 prevented an
unprivileged exploit on Pixel~9. These results demonstrate that kernel exploits can
generally be adapted to real Android devices.

\begin{table}[t]
  \centering
  \small
  \setlength{\tabcolsep}{6pt}
  \caption{Autonomous exploit adaptation outcomes across nine evaluated CVEs on commercial Google Pixel smartphones. A16 and A14 denote Android 16 and Android 14 releases; a dash (---) indicates the target was not in that bug's evaluated firmware subset.}
  \label{tab:physical-adaptation}
  \vspace{3pt}
  \begin{tabular}{lcccc}
    \toprule
    & \textbf{Pixel 10} & \multicolumn{2}{c}{\textbf{Pixel 9}} & \\
    \cmidrule(lr){2-2} \cmidrule(lr){3-4}
    \textbf{CVE} & \textbf{A16} & \textbf{A16} & \textbf{A14} & \textbf{Total} \\
    \midrule
    CVE-2024-26640 & ---  & ---  & Pass & 1/1 \\
    CVE-2024-38580 & ---  & ---  & Pass & 1/1 \\
    CVE-2024-46740 & ---  & ---  & Pass & 1/1 \\
    \addlinespace[2pt]
    CVE-2025-39977 & Pass & Fail & Fail & 1/3 \\
    \addlinespace[2pt]
    CVE-2026-43284 & Pass & ---  & ---  & 1/1 \\
    CVE-2026-43499 & Pass & Pass & ---  & 2/2 \\
    CVE-2026-53361 & Pass & Pass & ---  & 2/2 \\
    CVE-2026-53362 & Pass & ---  & ---  & 1/1 \\
    CVE-2026-64560 & Pass & Pass & ---  & 2/2 \\
    \midrule
    \textbf{Pass rate} & \textbf{6/6} & \textbf{3/4} & \textbf{3/4} & \textbf{12/14} \\
    \bottomrule
  \end{tabular}
\end{table}

\section{N-Day Exposure Analysis}
\label{sec:patch-propagation-results}

\begin{figure}[!t]
  \centering
  \includegraphics[width=0.47\textwidth]{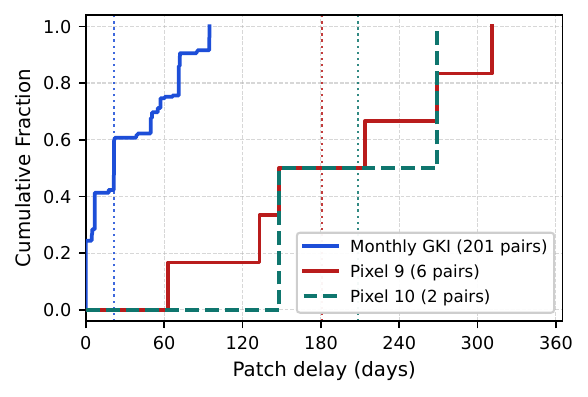}
  \caption{Cumulative distribution function (CDF) of patch propagation delays since upstream patch date, excluding unpatched CVE--branch pairs.}
  \label{fig:patch-propagation-cdf}
\end{figure}

Prior work on patch propagation delays quantifies how long fixes take to arrive~\cite{zhang2021androidpatch,fiber}, but does not, by itself, establish whether the underlying vulnerabilities are exploitable on Android. Our pipeline’s successful exploitation of 17 CVEs allows us to measure how long Android devices remain exposed to these n-day vulnerabilities that can be exploited in practice. For these CVEs, we quantify n-day exposure by measuring patch propagation delays and exposure lower bounds where fixes remain absent. We then analyze propagation patterns, branch maintenance, and security bulletin coverage to identify where remediation stalls and examine the factors behind prolonged exposure.

\subsection{Measuring N-Day Exposure}

In this section, we measure two exposure windows:

1. \emph{Post-patch exposure window}: elapsed time from upstream fixing commit to a downstream patch. This is effectively patch propagation delay~\cite{zhang2021androidpatch}. We rename it as an exposure window because of the heightened exploitation risks as demonstrated in our pipeline.

2. \emph{Post-disclosure exposure window}: elapsed time from public CVE
disclosure to a downstream patch. This is a more conservative estimate of the exposure window, assuming an attacker tracks only CVEs instead of individual patches or commits from upstream. CVEs are typically assigned days or weeks after the initial patch is made public.

Across the 197 monthly GKI release branches, 719 CVE--branch pairs were patched, but only 201 had an observable vulnerable interval and therefore contribute to our patched exposure measurements. The other 518 do not; for example, they were already patched at the time of forking.

There are 915 unpatched CVE--branch pairs; we can establish exposure bounds for 196: 114 on deprecated branches that remained unpatched when support ended after the upstream fix, measured through their deprecation dates, and 82 on active branches. We exclude the remaining 719 deprecated-branch pairs because 661 were deprecated before the upstream fix and 58 lack documented deprecation dates.

\PP{Post-patch exposure window (patched pairs).}
According to Appendix Table~\ref{tab:patch-propagation-horizons}, for the 201 pairs that were initially vulnerable on monthly GKI branches, patches propagated in a median of 21.5 days (mean 32.4 days, maximum 94.7 days). Over a third of these pairs (39.3\%, 79/201) remained unpatched after 30 days, and 25.4\% (51/201) remained unpatched after 60 days.
By comparison, upstream patches required a median of 181.0 days to reach Pixel firmware releases. Across the 8 patched CVE--device pairs, every patch required over 60 days to reach end users (100.0\%, 8/8), seven-eighths (87.5\%, 7/8) remained unfixed after 90 days, and half (50.0\%, 4/8) remained unfixed after 180 days (Appendix Table~\ref{tab:patch-propagation-horizons}).
We also visualize the delays in Figure~\ref{fig:patch-propagation-cdf}. As we can see, the patch-delay bottleneck is clearly in ACK-to-Pixel propagation rather than upstream-to-ACK propagation.

\PP{Post-patch exposure window (unpatched pairs).}
Across the 196 unpatched CVE--branch pairs, exposure exhibits a median lower bound of 146.8 days (mean 203.8 days, maximum 715.2 days), with 80.6\% (158/196) remaining unpatched beyond 90 days.
Combining the 201 patched pairs with all 196 measurable unpatched pairs yields a median exposure lower bound of 71.4 days across 397 pairs (mean 117.0 days), with at least 55.9\% (222/397) remaining unpatched after 60 days and at least 44.1\% (175/397) after 90 days.
Similarly, 9 evaluated CVE--device pairs remain open and unpatched across all evaluated Pixel firmware releases, exhibiting a median lower bound of 154.0 days (maximum 183.0 days).

\PP{Post-disclosure exposure window (patched + unpatched pairs).}
Using NVD CVE publication dates, 188 CVE–branch pairs exhibit positive post-disclosure exposure windows: 106 were patched after disclosure, with a median exposure of 20.9 days (maximum 64.1 days), while 82 remained unpatched at their branch tips, with a median lower-bound exposure of 86.4 days (maximum 640.3 days). This is a substantial reduction from the post-patch baseline: the number of patched pairs with positive exposure falls from 201 to 106, while the number of measurable unpatched pairs falls from 196 to 82. Accordingly, the overall population with positive exposure decreases from 397 pairs under the post-patch baseline to 188 under the post-disclosure baseline. 

On Pixel devices, 8 patched CVE--device pairs exhibit a median post-disclosure exposure of 149.5 days (maximum 262.0 days), while 9 pairs remain open and unpatched through the latest evaluated release, with a median lower bound of 94.0 days. Although measuring from CVE disclosure shortens the exposure window relative to the upstream-patch baseline, this reduction is small compared with the delay in propagating fixes to Pixel devices: the median for patched pairs decreases only from 181.0 to 149.5 days. Thus, although CVE disclosure typically follows the upstream patch by only a matter of weeks in our measurements, this lag does little to reduce end-user exposure; ACK-to-Pixel patch propagation remains the dominant source of delay.

\subsection{Understanding Prolonged N-Day Exposure}

To understand the factors behind prolonged n-day exposure, we first examine how branch maintenance affects patch propagation, then consider how gaps in security bulletin coverage may affect prioritization, and finally analyze delivery delays that persist after public demonstrations of exploitability.

\PP{Routine kernel maintenance does not ensure vulnerability remediation.}
We find that although ACK branches such as \texttt{android15-6.6} receive upstream fixes regularly over multiple years, devices use particular dated release GKI branches derived from them, such as \texttt{android15-6.6-2026-04}. These release branches receive selected fixes rather than continuously tracking the latest parent-branch code. A device can therefore receive kernel updates within the same release branch while missing fixes already available elsewhere in the parent or other derived ACK branches. Our measurements show this distinction in practice. For CVE-2026-31648, a fixed 6.6 release branch became available on July 1, 2026, yet Pixel 10 remained on an older unpatched branch through September 15, spanning 76 days, three firmware releases, and two kernel updates. In fact, among the eight patched CVE--device pairs, four received their fixes through a switch to another release branch while their previous branches remained unpatched. These switches occurred within the same kernel line, e.g., different derived branches from \texttt{android15-6.6}. Thus, continued support and regular kernel updates do not guarantee that a device receives a particular security fix.

\PP{Missing security signals may leave fixes to routine integration.}
CVE-2024-26640 and CVE-2024-38580 on Pixel 9, together with CVE-2025-39977 on both Pixel models, illustrate particularly long waits for remediation. The ACK commits carrying the fixes ultimately shipped in firmware were dated 304, 196, and 257 days before that delivery, respectively, with the latter interval applying to both devices. 
Even after ACK versions containing these fixes were tagged on the devices’ respective kernel lines, another 153--192 days passed before the fixes reached Pixel firmware.
In every case, remediation arrived through a branch switch while the previous branch remained unpatched. During this interval, four or five firmware releases shipped without the fixes, including releases that did not change the kernel revision. Notably, none of these three CVEs appeared in an ASB by the September 2026 cutoff, reflecting the limited ASB coverage of our exploited vulnerabilities overall: only three of 17 were listed at the time of writing (Oct 2026). These cases motivate a possible explanation for prolonged exposure: without an explicit signal to prioritize a security fix, remediation may depend on a broader branch update rather than a targeted backport to the deployed branch.

\PP{Public visibility does not guarantee rapid delivery.}
Publicized vulnerabilities exhibit shorter observed delivery delays among the CVE--device pairs that were patched. Across the three patched pairs involving GhostLock (CVE-2026-43499) on both Pixel models and Dirty Frag (CVE-2026-43284) on Pixel 9, fixes reached firmware a median of 96 days after ACK releases containing them became available for the corresponding devices’ kernel lines (range: 90--110 days), compared with 161 days for the other five patched pairs. This association is consistent with public attention helping prioritize remediation, although differences in branch adoption and the remaining unpatched cases prevent attributing it to publicity alone.
Nevertheless, GhostLock's explicit Android exploitation writeup still preceded fixed Pixel firmware by 62 days. Similarly, CVE-2026-64560 remained unpatched in the latest evaluated Pixel firmware 44 days after publication of an exploit explicitly demonstrated on Pixel 10 Pro.
Even where public Android exploit code was unavailable, Android relevance had not necessarily gone unrecognized. BadGarbage (CVE-2026-53361) and IPV6\_FRAG\_ESCAPE (CVE-2026-53362) were publicly discussed as relevant to mobile devices in July, yet both remained unpatched on the evaluated Pixel models through September 15. These cases distinguish an absence of public Android exploit code from an absence of public warning signals: Android relevance can already be public while remediation remains incomplete.

Our pipeline demonstrates that Android relevance can be identified automatically from public upstream patches and substantiated through validated exploitation on Android targets. Vendors therefore need not wait for bulletin inclusion or high-profile Android exploit releases to identify fixes requiring security attention. Relying on these delayed, selective signals and routine branch updates is no longer adequate for managing n-day exposure.

\section{Exploit Strategies and Techniques}
\label{sec:exploit-strategies}

In this section, we investigate the exploitation strategies and techniques that made the LLM-generated exploits successful. This sheds light on the exploitation playbook that may be represented in the models' training data.
Despite production mitigations---including KASLR, PAN, PXN, CFI, the \code{untrusted_app} SELinux domain, and seccomp filtering---the agents find viable paths from memory-safety bugs to local privilege escalation.
Table~\ref{tab:verified-strategies} categorizes these mechanisms across three dimensions: \emph{exploit strategy} (the overall exploitation path), \emph{exploit technique} (mechanisms for memory layout, timing, and primitive construction), and \emph{reliability measures} (a specific category of techniques which aim to improve fault tolerance).

We also analyze the exploits generated by our agents to extract their underlying strategies and techniques and distinguish established approaches from behaviors that appear novel. OpenAI recently reported that frontier models can develop novel exploitation techniques in Linux kernel exploits~\cite{openai_lpe}, but released no technical details of these behaviors. Android presents a substantially more constrained exploitation environment, rendering many conventional Linux strategies and techniques inapplicable. Nevertheless, we independently observe similar capabilities in our Android experiments. We document these behaviors here, providing concrete examples of novel exploit techniques generated under Android's tighter security constraints.

\subsection{Known Strategies and Techniques}

From Table~\ref{tab:verified-strategies}, we can see that the model is familiar with a large variety of common exploit strategies and techniques. We summarize a few of them below.

\PP{Constrained read/write versus arbitrary read/write.}
We classify each verified exploit by the strongest write primitive it constructs, separating control of the destination from control of the written value. Ten of the 17 exploits obtain an arbitrary-value, arbitrary-address write; four obtain an arbitrary-address write with a constrained value, such as a decrement; and three can write arbitrary values only to constrained addresses. For example, in CVE-2026-80778 the agent directs an arbitrary address decrement first at \code{selinux_state.enforcing}, changing it from one to zero, and then at the process's all-zero capability mask, underflowing it to all ones before calling \code{setresuid(0, 0, 0)}. Thus, 14 exploits control the write address, 13 control the written value, and 10 control both. Arbitrary-address reads are less common: five exploits obtain one, nine obtain constrained reads, and three complete exploitation without a read primitive.

\begin{table}[!t]
  \centering
  \small
  \caption{Recurring mechanisms across the 17 verified Android kernel exploits, categorized by exploitation strategy, exploit technique, and reliability measure.}
  \label{tab:verified-strategies}
  \begin{tabularx}{\columnwidth}{@{}Xr@{}}
    \toprule
    \textbf{Category / Mechanism} & \textbf{Cases} \\
    \midrule
    \multicolumn{2}{@{}l}{\textbf{Exploit Strategy (Overall Exploitation Path)}} \\
    Direct credential field edit or capability underflow~\cite{badnode_writeup} & 7 \\
    Task credential pointer replacement~\cite{dirtycred,califioCVE202538617} & 6 \\
    Page-table corruption or forged hierarchy~\cite{projectzeroKernelMemory} & 5 \\
    Shared executable page-cache modification~\cite{dirtypipe} & 3 \\
    CFI compatible function pointer replacement & 1 \\
    Address leak via debug-interface after disabling SELinux & 5 \\
    \midrule
    \multicolumn{2}{@{}l}{\textbf{Exploit Technique (Memory Layout, Timing, \& Primitives)}} \\
    CPU placement, affinity, or core pinning~\cite{kheaps} & 16 \\
    Socket or network object used for grooming or reuse~\cite{lexfoCVE201711176} & 10 \\
    Cross-cache reuse through the buddy allocator~\cite{slubstick} & 8 \\
    Same-cache object substitution~\cite{slake} & 7 \\
    Kernel decrement redirected to chosen target address~\cite{nccTriforceAFL} & 7 \\
    Lock contention or deliberately blocked operations~\cite{califioCVE202538617} & 6 \\
    Tree or linked-list maintenance shaped into pointer store~\cite{badnode_writeup} & 5 \\
    Timer-induced race window expansion~\cite{projectzeroRaceClock} & 4 \\
    CPU saturation or worker starvation~\cite{kernelctfCVE202450264} & 4 \\
    \midrule
    \multicolumn{2}{@{}l}{\textbf{Reliability Measure (Runtime Checks \& Fault Tolerance)}} \\
    Application-visible progress check~\cite{projectzeroRaceClock} & 17 \\
    Disposable worker process, watchdog, or retry limit~\cite{badepoll} & 16 \\
    Fault-tolerant memory encoding~\cite{kernelctfCVE20234623} & 1 \\
    Post-exploit state repair~\cite{lexfoKernelRepair} & 3 \\
    Syscall latency side channel to detect race win & 1 \\
    \bottomrule
  \end{tabularx}
\end{table}

\PP{Memory grooming and allocator shaping.}
To groom the heap and manipulate timing (Table~\ref{tab:verified-strategies}, middle), the agents rely predominantly on socket structures (\code{sk_buff} headers, TCP objects, and \code{AF_UNIX} datagrams) that unprivileged apps can allocate in volume without special permissions.
When slab cache isolation blocks direct same-cache substitution, the agents perform cross-cache reclamation~\cite{cross-x}: freeing objects in bulk releases slab pages back to the buddy/page allocator, allowing them to be reacquired as page tables, DMA-buf heaps, or pipe buffers.
CPU core pinning stabilizes these layouts by preventing inter-core slab migrations.

\PP{Failure isolation and runtime verification.}
To tolerate nondeterminism (Table~\ref{tab:verified-strategies}, bottom), the agents isolate volatile steps in disposable child processes and gate exploitation steps with application-visible progress checks rather than sleeping and assuming success.
Four exploits construct fault-tolerant memory patterns or repair corrupted state post-exploitation to prevent watchdog panics.
Several techniques also improve reliability: CPU pinning prevents thread migration across cores, timers widen race windows, and lock contention or socket stalls keep memory allocations stable during races.

\subsection{Novel Exploitation Strategies and Techniques}
\label{sec:novel-techniques}

Beyond established mechanisms, the agents introduced two novel strategies and three novel techniques:

\PP{Object address leak via debug interfaces once SELinux is disabled.}
Android prevents unprivileged apps from inspecting kernel symbols, \code{/proc/pagemap}, or tracepoints.
Rather than finding an address leak beforehand, several exploits invert this sequence: they disable security enforcement first, then use debugging interfaces to locate targets.
In CVE-2026-80778, the bug provides an atomic decrement without an address leak.
The agent decrements a fixed kernel image offset to clear \code{selinux_state.enforcing} (KASLR is bypassed on Pixel and Cuttlefish~\cite{projectzeroDefeatKASLR}).
Disabling SELinux grants access to debugging interfaces, including tracefs tracepoints (\code{kmem/}\allowbreak\code{kmem_}\allowbreak\code{cache_}\allowbreak\code{alloc}) and perf events via Perfetto.
The agent spawns a child while tracing, reads the child's \code{struct cred} kernel address from the trace, and directs its second decrement at the capability mask.
Four other exploits similarly disable or corrupt SELinux first to locate target objects.

\PP{CFI-compatible LSM function address rewrite strategy.}
Android kernels enforce CFI, preventing indirect calls to arbitrary addresses.
In CVE-2026-46215, rather than bypassing CFI or altering credentials, the agent redirected the LSM static-call key \code{__SCK__}\allowbreak\code{lsm_}\allowbreak\code{static_}\allowbreak\code{call_}\allowbreak\code{capable_0} to \code{safesetid_}\allowbreak\code{security_}\allowbreak\code{capable}.
Because this function shares the capability hook prototype, it passes CFI checks while returning success for in-set-ID checks, allowing an unprivileged \code{setresuid(0, 0, 0)} call to grant root. This means that assuming KASLR is bypassed, a single arbitrary write primitive corrupting the static key can successfully lead to root.

\PP{msg\_control buffers with fully controllable lifetime and data}
In CVE-2026-46215, the agent reclaims a freed 512-byte DRM GEM object using socket control message buffers (\code{msg_control}).
Prior control-buffer sprays persist the allocation by blocking \code{sendmsg()}, but require a valid \code{cmsghdr} which constrains the first 12 bytes of the object ~\cite{eshardKernelUAFWrite,githubAndroidMitigationsRace}.
The agent avoids this restriction by using \code{MSG_FASTOPEN} with a server whose accept queue is full. The connection remains pending, so the kernel keeps the copied control buffer allocated even though its header is invalid. This gives the agent full control over the contents and lifetime of an elastic object (in this case a 512-byte object in \code{kmalloc-512}).

\PP{Inode lock contention for heap-buffer lifetime extension.}
On Android, unprivileged apps cannot rely on the conventional \code{userfaultfd} or FUSE techniques to stall kernel-mode page faults during heap spraying. To retain long-lived, fully attacker-controlled heap payload allocations, the agent exploiting CVE-2024-38580 induces file-lock contention during \code{setxattr()} calls.
Because \code{setxattr_copy()} allocates the attacker-controlled buffer before the inode lock is acquired, blocked setter threads retain their buffers and keep fake DMA-BUF objects stable throughout the race window.
The closest related technique we are aware of is the ``fallocate hole-punch'' technique which uses filesystem synchronization to prolong kernel object lifetimes~\cite{fallocate_hole_punch}, but the mechanism of stalling is blocking the shmem page fault instead.
Interestingly, the \code{msg_control} object mentioned above would replace this object perfectly with cleaner lifetime control, but this agent did not discover it.

\PP{Syscall-timing side channel for race success detection when a WARN is triggered.}
In CVE-2026-64560, winning the race triggers an internal kernel \code{WARN_ON}.
Because Android's \code{untrusted_app} policy blocks access to \code{dmesg}, the agent cannot observe this warning directly.
Instead, it measures syscall latency: kernel formatting overhead causes a 10\,ms spike versus normal 1\,ms execution, providing a non-destructive runtime oracle that confirms race victory before initiating memory reclamation.

\section{Discussion}
\label{sec:discussion}

\PP{Agent awareness and nondeterminism.}
As described in Section~\ref{sec:exploit-outcomes}, three evaluated vulnerabilities with published exploits exhausted their budget before passing blind verification.
Because exploitation of these bugs is known to be possible, these results suggest that the agent may lack awareness of race or other kernel exploit techniques.
However, agent decision-making is nondeterministic, and re-running experiments may explore alternative strategies and yield different outcomes or additional successes.

\PP{Prompt design.}
Our prompts only describe the environment, available tools, and objective.
We provide no exploitation guidance, templates, kernel documentation, or domain-specific knowledge.
This forces the agent to rediscover viable exploitation strategies and techniques from scratch, unless they are already prominent in its training data.
We suspect that supplying domain knowledge regarding techniques or strategies that required extensive search or were overlooked entirely can steer the agent away from dead ends and reduce exploration cost.

\PP{Exploitability lower bound.}
The 17 verified exploits represent a lower bound on the true exploitability of the evaluated targets.
Most experiments terminated due to exhausting their fixed budget, and several attempts had achieved LPE before running out of budget during stabilization.
Higher or dynamically allocated budgets would likely yield additional verified exploits.
Despite being a lower bound, these exploits span the entire evaluated lifespans of Google Pixel~9 (August 2024--October 2026) and Pixel~10 (August 2025--October 2026).
An attacker therefore does not require an exhaustive or expensive exploit pipeline to sustain unbroken access across device generations.

\PP{Implication on defenders.}
Our results show that existing patch-downstreaming and vulnerability-disclosure processes are neither fast nor comprehensive enough to keep pace with automated exploit development. Defenders can no longer assume that a public fix will not be used to weaponize a bug before downstreaming.
While we approach this from the position of an attacker, this tooling is dual purpose. Defenders can use the same pipeline to screen patches, reproduce bugs, and validate backports in order to operate on a shorter timeline.
Furthermore, downstream maintainers (e.g., Google) can use a specialized pipeline to prioritize bug reports before a patch is available, enabling them to quickly deploy mitigations.

\section{Related Work}
\label{sec:related-work}

\PP{Android and Linux Patch Propagation}
Prior work measures how security fixes propagate through the fragmented Android
and Linux ecosystems. Zhang et al.~\cite{zhang2021androidpatch} track security
patches across upstream Linux, the Android Common Kernel (ACK), and vendor
kernels, showing that fixes frequently take months or years to reach devices.
Li et al.~\cite{li2024investigation} analyze downstream Linux distributions and
characterize delays, omissions, and inheritance patterns across branches. Maar
et al. examine unpatched n-day exposure in vendor drivers~\cite{maar2025doom} and
study the deployment and effectiveness of defenses against public Android
exploit techniques~\cite{maar2024defects}. We complement these ecosystem studies
by measuring upstream fix propagation across individual ACK branches and Pixel
firmware releases, and we compare these exposure horizons directly to the
development time required by an autonomous exploit-generation pipeline.

\PP{Automated Vulnerability Analysis and Reproduction}
Automated kernel bug reproduction systems bridge the gap between public reports
and downstream execution environments. SyzBridge~\cite{syzbridge} adapts
upstream syzbot reproducers to downstream kernel distributions, and
AEM~\cite{jiang2023aem} evaluates exploit migration across kernel versions.
KernJC~\cite{kernjc} reconstructs vulnerable kernel environments from CVE
metadata, while SyzScope~\cite{syzscope} and GREBE~\cite{lin2022grebe} identify
stronger security consequences or alternative triggering contexts for known
bugs. Recent work explores using LLM agents for bug reproduction:
Patch-to-PoC~\cite{pu2026patch} evaluates whether LLM agents can reproduce Linux
kernel n-days from fixing commits in prepared kernelCTF environments.
CyberGym~\cite{wang2026cybergym} is a benchmark that evaluates LLM agents for bug reproduction in various user-space programs. Our work
shares the goal of from-patch reproduction but integrates discovery into an
end-to-end pipeline that selects downstream Android targets, develops
reproducers without human assistance, and continues through full exploit
generation and verification.

\PP{Automated Exploit Generation}
Traditional kernel automated exploit generation (AEG) systems automate specific
phases of exploitation. FUZE~\cite{fuze} evaluates exploitability and
synthesizes kernel exploits for use-after-free bugs, KOOBE~\cite{koobe} analyzes
out-of-bounds write exploitation, and Revery~\cite{revery} searches for usable
exploitation paths. K-LEAK~\cite{kleak} automates multi-step information leaks,
AlphaEXP~\cite{wang2023alphaexp} identifies security-sensitive kernel objects,
and related studies systematize memory layout manipulation via page spraying and
SLUBStick~\cite{pagespray,slubstick}. These systems implement specialized
symbolic execution or domain-specific heuristics tailored to individual bug
classes. In contrast, LLM agents reason over source code, commit history,
and runtime execution feedback across diverse vulnerability classes.

Recent benchmarks evaluate the offensive capabilities of LLM agents.
PwnGPT~\cite{peng2025pwngpt} evaluates LLM-assisted exploit
generation on synthetic CTF binaries.
ExploitGym~\cite{wang2026exploitgymaiagentsturn} is a
benchmark for LLM-driven exploit generation, evaluating agents across userspace
programs, browser engines, and Linux kernels. We adopt its secret-retrieval
success criterion and LLM-based review for vulnerability attribution. Our pipeline continuously
evaluates the broader path from a public upstream fix to a verified Android
exploit and relates exploit-development time to downstream patch propagation.

\section{Conclusion}
\label{sec:conclusion}

Autonomous exploit generation fundamentally shifts the threat model of public vulnerability disclosure.
What previously required specialized kernel expertise and substantial manual effort can now be executed by LLM agents for hundreds of dollars per exploit.
This capability makes downstream patch propagation delays a critical liability. Defending mobile ecosystems requires treating public upstream commits as immediate exploitation signals and prioritizing rapid downstream patch distribution.

\clearpage

\bibliographystyle{IEEEtran}
\bibliography{reference}

\clearpage
\appendices
\section{Supplementary Pipeline Results}
\label{app:supplementary-pipeline-results}

Table~\ref{tab:patch-propagation-horizons} summarizes post-patch exposure for
initially vulnerable monthly GKI branches and patched Pixel device pairs.
Table~\ref{tab:reliability} reports fresh-instance reliability for the 17
verified exploits, including confidence intervals and the CVEs in each
pass-rate group.

\begin{table}[!htbp]
  \centering
  \footnotesize
  \setlength{\tabcolsep}{1.8pt}
  \caption{Post-patch n-day exposure window (patch propagation delay) and percentage of CVE--target pairs remaining unpatched after 30, 60, 90, 180, and 365 days from the upstream fixing commit.}
  \label{tab:patch-propagation-horizons}
  \vspace{3pt}
  \begin{tabular*}{\columnwidth}{@{\extracolsep{\fill}}lrrrrrrr@{}}
    \toprule
    & & & \multicolumn{5}{c}{\textbf{Unpatched after}} \\
    \cmidrule(lr){4-8}
    \textbf{Target} & \textbf{Pairs} & \textbf{Median} & \textbf{30d} & \textbf{60d} & \textbf{90d} & \textbf{180d} & \textbf{1yr} \\
    \midrule
    \makecell[l]{Monthly GKI\\(initially vuln.)} & 201 & 21.5d & 39.3\% & 25.4\% & 8.5\% & 0.0\% & 0.0\% \\
    Pixel releases & 8 & 181.0d & 100.0\% & 100.0\% & 87.5\% & 50.0\% & 0.0\% \\
    \quad Pixel 9 & 6 & 181.0d & 100.0\% & 100.0\% & 83.3\% & 50.0\% & 0.0\% \\
    \quad Pixel 10 & 2 & 208.5d & 100.0\% & 100.0\% & 100.0\% & 50.0\% & 0.0\% \\
    \bottomrule
  \end{tabular*}
\end{table}
\begin{table}[t]
  \centering
  \small
  \setlength{\tabcolsep}{4pt}
  \caption{Fresh-instance execution reliability across 17 verified exploits (ten trials each). Wilson 95\% confidence intervals are reported in brackets.
    }
  \label{tab:reliability}
  \begin{tabularx}{\columnwidth}{@{}lrr>{\raggedright\arraybackslash}X@{}}
    \toprule
    \textbf{Pass rate (95\% CI)} & \textbf{Cases} & \textbf{Share} & \textbf{Vulnerabilities} \\
    \midrule
    10/10 [72.2, 100]\% & 7 & 41.2\% & CVE-2024-26640, CVE-2024-46740, CVE-2024-56554, CVE-2025-39977, CVE-2026-43284, CVE-2026-46215, CVE-2026-53362 \\
    \addlinespace[2pt]
    9/10 [59.6, 98.2]\% & 5 & 29.4\% & CVE-2024-38580, CVE-2026-43499, CVE-2026-52973, CVE-2026-53361, CVE-2026-80778 \\
    \addlinespace[2pt]
    8/10 [49.0, 94.3]\% & 3 & 17.6\% & CVE-2024-36972, CVE-2026-23077, CVE-2026-31648 \\
    \addlinespace[2pt]
    3/10 [10.8, 60.3]\% & 1 & 5.9\% & CVE-2024-36971 \\
    \addlinespace[2pt]
    1/10 [1.8, 40.4]\%  & 1 & 5.9\% & CVE-2026-64560 \\
    \bottomrule
  \end{tabularx}
\end{table}

\newpage

\section{Cross-Target Exploit Adaptation Breakdown}
\label{app:adaptation-breakdown}

Table~\ref{tab:adaptation-costs} details the model costs and primary technical challenges encountered across all 34 same-branch and adjacent-branch adaptation experiments for the 17 verified CVEs.

\begin{table*}[!t]
  \centering
  \small
  \caption{Exploit adaptation costs and technical challenges across same-branch (newer commit) and adjacent-branch kernel targets for all 17 verified vulnerabilities. Costs denote API-equivalent model expenditures after subtracting starting expenditure.}
  \label{tab:adaptation-costs}
  \begin{tabularx}{\textwidth}{@{}lrrX@{}}
    \toprule
    \textbf{CVE} & \textbf{Same Branch} & \textbf{Adjacent Branch} & \textbf{Primary Technical Challenge} \\
    \midrule
    CVE-2024-26640 & \$6.54 & \$8.08 & Kernel 5.15 shifted \code{task_struct} offsets; both required constant/layout retargeting only. \\
    CVE-2024-36971 & \$512.72 & \$411.29 & Same-branch THP allocation and route timing drifted; 5.15 SLUB partial-list semantics forced a redesigned memory reclamation campaign. \\
    CVE-2024-36972 & \$16.94 & \$1.88 & Adjacent transferred without source changes; same-branch required safe post-exploitation object cleanup. \\
    CVE-2024-38580 & \$4.76 & \$292.34 & Kernel 5.15 lacked DMA-BUF sync-file ioctls, forcing synthesis of a new composite epoll, Binder, and DMA-BUF payload object. \\
    CVE-2024-46740 & \$12.34 & \$38.74 & Kernel 6.6 Binder KABI expansion shifted the target object to an adjacent slab cache. \\
    CVE-2024-56554 & \$4.41 & \$431.04 & Kernel 6.1 Binder cache geometry broke the double-free sequence, requiring a reconstructed freelist corruption and TCP write chain. \\
    CVE-2025-39977 & \$87.47 & \$838.09 & Kernel 6.6 lacked the 6.12 EEVDF scheduler fields, requiring a complete redesign of the fake task and write gadget. \\
    CVE-2026-23077 & \$11.59 & \$32.51 & Both plan entries targeted an identical kernel; cost divergence reflects variation in LLM exploration. \\
    CVE-2026-31648 & \$7.85 & \$20.54 & Both targets shifted task offsets; the adjacent target also encountered Android process limits. \\
    CVE-2026-43284 & \$1.90 & \$1.61 & Both APKs adapted without exploit-code modifications. \\
    CVE-2026-43499 & \$156.04 & \$368.91 & Scheduler and priority-inheritance ABI drift required re-engineering fake-task placement in kernel memory. \\
    CVE-2026-46215 & \$8.81 & \$9.61 & Both preserved the DRM GEM primitive and required address mapping and device-layout updates only. \\
    CVE-2026-52973 & \$9.71 & \$9.03 & Primitives and allocator layouts survived; required tuning tracepoint identifiers and race timing. \\
    CVE-2026-53361 & \$66.63 & \$117.13 & Disabled \code{AF_UNIX} out-of-band data required an alternate payload object; adjacent also rebuilt a transient UAF chain. \\
    CVE-2026-53362 & \$5.59 & \$5.90 & KABI changes relocated \code{sk_buff.users}, but allocator geometry and the vulnerable IPv6 path remained intact. \\
    CVE-2026-64560 & \$148.40 & \$488.99 & Kernel 6.18 introduced timer, credential, and LSM layout drift, forcing redesign of KASLR bypass and heap reclamation. \\
    CVE-2026-80778 & \$35.78 & \$29.91 & HW-tag KASAN altered allocation strides; same-branch required a credential leak while adjacent retuned signal-object spraying. \\
    \midrule
    \textbf{Total / Median} & \textbf{\$1,097.49 / \$11.59} & \textbf{\$3,105.59 / \$32.51} & \textbf{All 34 adaptations passed independent blind verification.} \\
    \bottomrule
  \end{tabularx}
\end{table*}

\end{document}